\documentclass[aps,prd,twocolumn,nofootinbib,superscriptaddress]{revtex4-2}
\usepackage[T1]{fontenc}
\usepackage[utf8]{inputenc}
\usepackage{amsmath,amssymb,amsthm}
\usepackage{graphicx}
\usepackage{hyperref}

\begin{document}

\title{Calling the Brane Next Door:\\[4pt]
\large The Kaluza--Klein Tower as a Gravitational Information Channel}

\author{Karim Benakli}

\affiliation{Sorbonne Universit\'e, CNRS, Laboratoire de Physique
Th\'eorique et Hautes \'Energies, LPTHE, F-75005 Paris, France}

\email{kbenakli@lpthe.jussieu.fr}
\date{June 8, 2026}

\begin{abstract}
Could two worlds localised on neighbouring branes communicate through
gravity alone? We investigate this question in a minimal
higher-dimensional framework in which Standard Model fields are
confined to our brane while gravity propagates through the bulk.
Starting from the brane-to-brane graviton propagator, we derive the
retarded transfer kernel of the inter-brane link and characterise the
transition from evanescent to propagating Kaluza--Klein modes.

The central idea is to give the Kaluza--Klein tower a new role: not
only as a spectrum of massive gravitational states, but as a set of
communication carriers. Below the first KK threshold, the channel is
effectively four-dimensional and is mediated only by the massless
graviton. Above threshold, massive KK modes open as additional
propagating subchannels. Information may then be encoded not only in
ordinary signal variables such as amplitude, phase, frequency, or
polarisation, but also in the occupation pattern, relative phases, and
arrival-time structure of the KK tower.

We formulate the resulting inter-brane link as a geometry-defined
linear channel. The compactification determines the KK masses,
wavefunctions, brane overlap factors, propagation phases, and
ultraviolet form factors; after projection onto transmitter and
receiver degrees of freedom, these data define 
a multi-input multi-output (MIMO) channel matrix.
In the resolved-mode limit, the tower yields a set of approximate
parallel subchannels, allowing standard information-theoretic notions
such as capacity bounds, water-filling, effective rank, and sparse
occupancy codes to be applied.

The practical production and detection of such signals are highly
model-dependent and are not assumed to be technologically feasible.
Nevertheless, the structure of the channel is well defined. A
neighbouring brane-world could be separated from us by a microscopic
distance in the compact space while remaining effectively hidden
because the only shared interaction is gravity. In such a scenario,
the first observable signature may not be a deliberate message, but
the spectral and modal structure of the Kaluza--Klein tower itself,
revealing partial information about the geometry of a nearby hidden
world.
\end{abstract}

\maketitle
\section{Introduction}
\label{sec:introduction}

Whether the Universe harbours intelligent life beyond our own world is
one of the oldest questions in natural philosophy. Yet the physics of
long-range communication is unforgiving: the nearest stars lie
light-years away, and most potentially habitable systems are orders of
magnitude more distant. The scenarios of interstellar contact 
that populate science fiction,
whether involving travel or communication, typically rely on
superluminal propagation, traversable wormholes, or exotic forms of
matter, all of which appear to be in tension with our present
understanding of fundamental physics.

In this paper we explore a different possibility. The proposal may
appear unusual at first sight, yet it invokes no violation of known
physical law. It rests on a simple geometrical observation: if our
Universe is a brane embedded in a higher-dimensional bulk, then another
world may be separated from us not by light-years of ordinary space,
but by a microscopic displacement in an extra dimension. The separation
between the two sectors may be microscopic, while the interaction
connecting them may be extremely weak. The relevant question is no
longer how to communicate across light-years using electromagnetic
radiation, but whether gravity alone can mediate information transfer
between neighbouring branes. In such a scenario, the nearest
civilisation need not orbit the nearest star; it could instead inhabit
a neighbouring brane.

The scenario rests on two assumptions. The first is the existence of a
second brane supporting localised degrees of freedom and, in principle,
observers. Such configurations arise naturally in string- and
M-theory-inspired constructions, where branes, boundaries, and
multi-brane geometries are standard ingredients. In most
phenomenological applications, hidden branes are treated as sectors
containing fields, particles, or sources of dark matter. The possibility
that they may harbour observers, and therefore be potential recipients
of information, has received comparatively little attention. This shift
of perspective transforms a standard problem of bulk field propagation
into a problem of inter-brane communication.

The second assumption is that gravity is the only mediator between the
two sectors. Additional bulk fields, such as scalars, axions, or dark
photons, may exist in particular models, but they are not universal.
The bulk graviton, by contrast, is present in any consistent
higher-dimensional theory. Focusing on gravity therefore isolates the
minimal channel that must exist if the brane-world picture is realised.
It also makes the problem maximally conservative: any additional bulk
field would only enlarge the set of possible communication channels.

The idea that spacetime may contain additional dimensions has a long
history, from the original Kaluza--Klein construction to modern
brane-world, large-volume, warped, and dark-dimension scenarios
\cite{Kaluza:1921tu,Klein:1926tv,Klein:1926fj,Einstein:1938fk,Fayet:1985kt,
Antoniadis:1990ew,Horava:1996ma,Arkani-Hamed:1998jmv,
Dienes:1998vh,Benakli:1998pw,Randall:1999ee,Montero:2022prj}.
These frameworks provide settings in which matter may be localised on
lower-dimensional defects while gravity propagates through the
higher-dimensional bulk. We do not commit to any specific
compactification. Our purpose is not to derive new constraints on a
particular model, but to isolate a question that can be posed in many
such constructions: if two sectors are localised on distinct branes and
gravity is the only field connecting them, what is the structure of the
gravitational communication channel?

The weakness of gravity is the central obstacle. In conventional SETI,
electromagnetic interactions are strong but the distances are vast. In
the brane-world setting considered here, the separation may be tiny,
but the shared interaction is Planck-suppressed. The possibility of
communication therefore depends not only on whether a disturbance can
propagate through the bulk, but on whether the higher-dimensional
gravitational sector provides additional structure that can be used as
a communication resource.

That structure is the Kaluza--Klein tower. Below the first KK
threshold, long-distance gravitational propagation is effectively
four-dimensional and is carried by the massless graviton. Above
threshold, massive KK modes become propagating. The channel then
acquires an internal mode structure absent in ordinary four-dimensional
gravity. Information need not be encoded only in the amplitude, phase,
frequency, or polarisation of a single waveform; it may also be encoded
in the distribution of excitation across KK levels, in their relative
phases, or in the mode-dependent arrival pattern produced by their
different group velocities.

This observation naturally leads to an information-theoretic
formulation. The compactification geometry determines the KK masses,
wavefunctions, brane overlap factors, and propagation phases. After
projection onto source and detector degrees of freedom, these data
define a linear channel matrix. In favourable regimes, the tower can be
treated approximately as a set of resolved subchannels; in more general
situations, it defines a multi-input multi-output (MIMO) channel whose optimal
communication basis is determined jointly by geometry and detector
noise. The extra dimensions therefore do more than modify the
gravitational spectrum: they define the signal space available for
communication.

The aim of this work is twofold. First, we characterise the physical
brane-to-brane gravitational channel in a minimal higher-dimensional
setup. Starting from the linearised bulk graviton propagator, we derive
the retarded propagation kernel, identify the KK thresholds, and
distinguish propagating from evanescent modes. Second, we formulate the
corresponding communication problem in information-theoretic terms. We
show how the physical Green function is projected onto a channel
matrix, discuss the conditions under which KK modes can be treated as
resolved subchannels, and analyse possible encoding strategies based on
KK-mode amplitudes and occupations.

Our objective is not to propose a realistic communication technology.
The production and detection of KK gravitons are highly
model-dependent and may require resources far beyond any foreseeable
capability. We instead ask a structural question: if a source on one
brane can populate the bulk gravitational channel, and if a receiver on
another brane can detect some projection of the resulting field, what
communication resources are supplied by the KK tower?

\medskip
\noindent\textbf{Scope of the present work.}
The primary objective of this paper is to characterise the physical and
information-theoretic structure of the inter-brane gravitational
channel. We determine the brane-to-brane propagator, analyse its
retarded finite-distance form, identify the KK thresholds, and discuss
how the resulting spectrum affects signal propagation, power transfer,
channel capacity, and encoding.

A separate question concerns the efficiency with which a given source
populates the KK tower. For any specific source class, the distribution
of emitted power among the graviton zero mode and the massive KK modes
depends on the compactification, the source stress tensor, greybody
factors, brane and bulk degrees of freedom, and detector response.
Determining these emission and detection rates requires a dedicated
analysis for each source model. We do not attempt such a calculation
here.

Accordingly, the discussion of candidate transmitters in
Section~\ref{sec:emitters} is illustrative rather than exhaustive. Its
purpose is to distinguish source regimes: high-power but low-frequency
compact mergers, broadband thermal black-hole emission, idealised
black-hole burst sources, and coherent but extremely weak
photon--graviton conversion. These examples clarify the trade-off
between power, bandwidth, and control; they are not engineering
proposals.

The paper is organised as follows. Section~\ref{sec:framework}
introduces the minimal brane-world setup and derives the brane-to-brane
graviton propagator, including the KK spectrum, overlap factors,
canonical coupling, and ultraviolet form factor. Section~\ref{sec:channel}
derives the retarded finite-distance propagation kernel and
distinguishes propagating from evanescent KK modes.
Section~\ref{sec:power} discusses power transfer through the bulk and
the difference between coherent and incoherent use of the tower.
Section~\ref{sec:emitters} compares candidate transmitter regimes.
Section~\ref{sec:signal} explains how the physical propagation kernel
is projected onto channel variables. Section~\ref{sec:capacity}
formulates the resulting information-theoretic channel and gives the
corresponding Gaussian capacity bound in the resolved KK limit.
Section~\ref{sec:encoding} discusses encoding in KK space, including
optimal communication bases, water-filling, sparse occupancy codes, and
the relation between channel tomography and inverse spectral geometry.
Section~\ref{sec:conclusions} summarises the results and limitations.



\section{A Minimal Framework for Inter-Brane Communication}
\label{sec:framework}

We consider the simplest setup capable of supporting communication
between two otherwise isolated worlds. The higher-dimensional
spacetime is a $(4+d)$-dimensional bulk containing two parallel
branes, $\mathcal{B}_A$ and $\mathcal{B}_B$, located at distinct
positions in the compact dimensions. Standard Model fields are
confined to $\mathcal{B}_A$. The matter content of $\mathcal{B}_B$
is left unspecified; we require only that it supports localised
observers capable of detecting gravitational signals. Gravity
propagates freely throughout the bulk. Throughout this section 
we use natural units, \(c=\hbar=1\), unless
otherwise stated.

The assumption of a second brane should not be understood as an
arbitrary addition. In string- and M-theory-inspired
compactifications, multi-brane configurations and boundary branes
arise naturally. In the Horava--Witten compactification of M-theory
on $S^1/\mathbb{Z}_2$~\cite{Horava:1996ma}, for example, the two
boundaries of the orbifold interval each carry an $E_8$ gauge
sector. We do not commit to this specific realisation, nor to any
particular microscopic construction. The point is only that a
second boundary with its own local degrees of freedom is a standard
possibility in ultraviolet completions of higher-dimensional
gravity.

The central idea explored in this work, namely the use of the
Kaluza--Klein tower as a communication resource, is already present
in the simplest case of a single compact extra dimension. We
therefore restrict our explicit analysis to a flat interval of
length $L$, with the two branes located at the endpoints,
\begin{equation}
  y_A = 0, \qquad y_B = L .
  \label{eq:brane_positions}
\end{equation}
More complicated compactifications modify the detailed spectrum,
wavefunctions, and overlap factors. The basic mechanism is already
fully visible in this minimal setting.

\begin{figure*}[t]
  \centering
  \includegraphics[width=\textwidth]{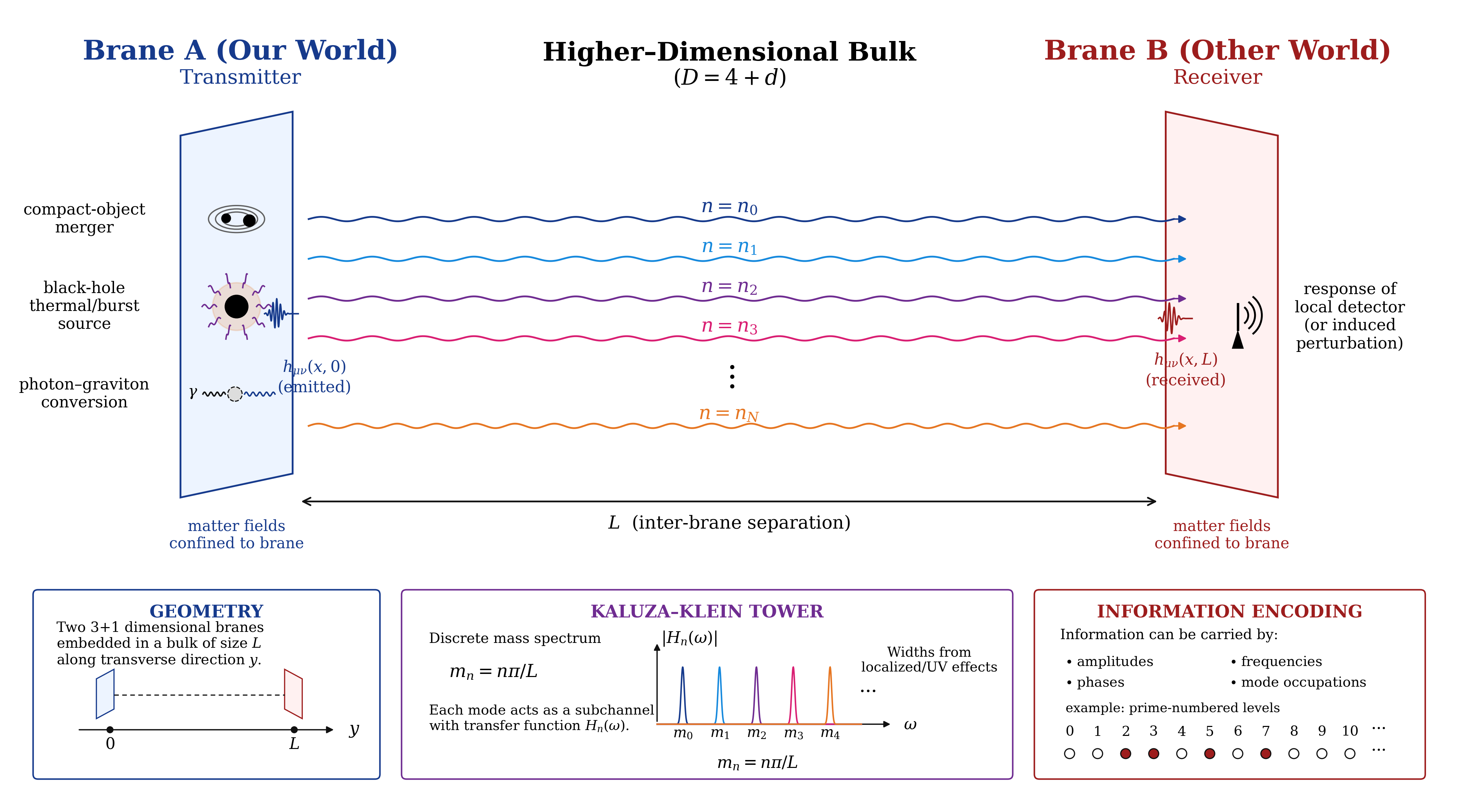}
 \caption{
Schematic representation of the inter-world gravitational communication
channel. Matter fields are localised on two distinct branes, while gravity
propagates through the higher-dimensional bulk. A source on brane
$\mathcal{B}_A$ excites the massless graviton and, at sufficiently high
frequency, massive Kaluza--Klein modes. These modes propagate through
the bulk and induce a response on brane $\mathcal{B}_B$. The KK tower
acts as a set of gravitationally coupled subchannels: information may be
encoded not only in the amplitude, phase, or frequency of a signal, but
also in the occupation pattern of KK levels.  }
  \label{fig:interbrane-channel}
\end{figure*}

The bulk metric is expanded around a flat background,
\begin{equation}
  g_{MN} = \eta_{MN} + \kappa_D\, h_{MN},
  \label{eq:metric_expansion}
\end{equation}
where $M,N=0,\ldots,3+d$, and $\kappa_D$ is the
$(4+d)$-dimensional gravitational coupling. Its precise normalisation
will not play a role below, since the subsequent discussion is
organised in terms of the canonically normalised four-dimensional KK
modes.

At linear order, matter localised on a brane couples to the induced
metric perturbation through its energy-momentum tensor. Restricting
the brane-intrinsic indices to $\mu,\nu=0,\ldots,3$, the interaction
action takes the schematic form
\begin{equation}
  S_{\rm int}
  =
  -\frac{\kappa_D}{2}
  \sum_{i=A,B}
  \int_{\mathcal{B}_i}
  d^4x\;
  h_{\mu\nu}(x,y_i)\,T_i^{\mu\nu}(x).
  \label{eq:SintD}
\end{equation}
This is the linearised coupling of the induced metric to brane matter.
In a complete model, finite brane thickness, stringy structure, or
localised kinetic terms can modify this coupling for sufficiently
massive KK modes; we address this in the discussion of the ultraviolet
form factor below.

A time-dependent source $T_A^{\mu\nu}$ on $\mathcal{B}_A$ excites
bulk gravitational degrees of freedom, which propagate through the
extra dimension and induce a response on $\mathcal{B}_B$. The linear
response is governed by the brane-to-brane graviton two-point
function,
\begin{equation}
  G^{\mu\nu\rho\sigma}_{AB}(x-x')
  =
  \bigl\langle
    h^{\mu\nu}(x,y_A)\,h^{\rho\sigma}(x',y_B)
  \bigr\rangle .
  \label{eq:GAB_tensor}
\end{equation}
Strictly speaking, $h_{\mu\nu}$ and its two-point function are
gauge-dependent quantities. Physical observables on the receiving
brane are gauge-invariant detector responses, such as tidal forces
or geodesic deviation. In this work we use the graviton propagator
in a fixed linearised gauge as a proxy for the channel transfer
function. The pole positions, KK thresholds, and mode-counting
statements are gauge-independent; the tensor numerator
$\Pi_n^{\mu\nu\rho\sigma}$ is gauge-dependent, and the detector
response is additionally model-dependent.

Expanding the bulk graviton in KK eigenmodes,
\begin{equation}
  h_{\mu\nu}(x,y)
  =
  \sum_n h^{(n)}_{\mu\nu}(x)\,f_n(y),
  \label{eq:KK_expansion}
\end{equation}
the brane-to-brane propagator in four-dimensional momentum space
takes the schematic form
\begin{equation}
  G^{\mu\nu\rho\sigma}_{AB}(p)
  =
  \sum_n
  f_n(y_A)\,f_n(y_B)\,
  \frac{\Pi_n^{\mu\nu\rho\sigma}(p)}
       {p^2-m_n^2+i\epsilon},
  \label{eq:GAB_tensor_KK}
\end{equation}
where $m_n$ are the KK masses and $\Pi_n^{\mu\nu\rho\sigma}$
denotes the spin projector and gauge-dependent numerator. The
massless zero mode and the massive KK gravitons carry different
tensor structures and different numbers of physical polarisations.
Since our main interest is the spectral organisation of the channel,
we suppress these tensor indices and write
\begin{equation}
  G_{AB}(p)
  =
  \sum_n
  f_n(y_A)\,f_n(y_B)\,
  \frac{1}{p^2-m_n^2+i\epsilon},
  \label{eq:GAB}
\end{equation}
where Eq.~\eqref{eq:GAB} represents the scalar part of the transfer
function in a fixed gauge.

For the flat interval with Neumann boundary conditions, the mode
functions are
\begin{equation}
  f_0(y)=\frac{1}{\sqrt{L}},
  \qquad
  f_n(y)=\sqrt{\frac{2}{L}}\,
  \cos\!\left(\frac{n\pi y}{L}\right),
  \quad n\geq1,
  \label{eq:flat_modes}
\end{equation}
with masses
\begin{equation}
  m_n=\frac{n\pi}{L}.
  \label{eq:KK_masses}
\end{equation}
The mode functions satisfy the orthonormality relation
\begin{equation}
  \int_0^L f_m(y)f_n(y)\,dy=\delta_{mn}.
\end{equation}
At the two endpoints,
\begin{equation}
  f_0(y_A)\,f_0(y_B) = \frac{1}{L},
  \quad
  f_n(y_A)\,f_n(y_B) = \frac{2}{L}(-1)^n,
  \quad n\geq1.
  \label{eq:endpoint_overlap}
\end{equation}
All levels of the flat-interval tower therefore have non-vanishing
support on both branes. In this idealised geometry the overlap
factors are mode-independent in magnitude, so the brane wavefunction
overlaps themselves do not suppress high KK levels.

It is important not to conflate the wavefunction normalisation in
Eq.~\eqref{eq:flat_modes} with the physical four-dimensional coupling
of canonically normalised KK gravitons. After dimensional reduction
and canonical normalisation of the four-dimensional fields, a resolved
KK graviton couples to brane matter with strength suppressed by the
reduced Planck mass,
\begin{equation}
  \bar M_{\rm Pl} \equiv (8\pi G_N)^{-1/2}
  \simeq 2.4\times 10^{18}~{\rm GeV}.
  \label{eq:reduced_planck_mass}
\end{equation}
For a flat interval with the brane at an endpoint, the interaction
Lagrangian takes the form
\begin{equation}
  \mathcal{L}_{\rm int}
  =
  -\frac{1}{\bar{M}_{\rm Pl}}\,h^{(0)}_{\mu\nu}T^{\mu\nu}
  -\frac{\sqrt{2}}{\bar{M}_{\rm Pl}}
  \sum_{n\geq1} h^{(n)}_{\mu\nu}T^{\mu\nu},
  \label{eq:KKcoupling_framework}
\end{equation}
up to convention-dependent factors at the level of the tensor
numerator. The factor $\sqrt{2}$ reflects the boundary value of the
normalised massive-mode wavefunction and is an $O(1)$ coefficient.
A single resolved KK mode is therefore Planck-suppressed in the same
parametric sense as the zero mode.

The multiplicity of the tower enters only in inclusive processes. Let
\(g_n\) denote the effective four-dimensional coupling of the \(n\)-th
canonically normalised KK graviton to brane matter. For a flat interval
with a brane at an endpoint, Eq.~\eqref{eq:KKcoupling_framework}
implies \(g_0\sim \bar M_{\rm Pl}^{-1}\) and
\(g_n\sim \sqrt{2}\,\bar M_{\rm Pl}^{-1}\) for \(n\geq 1\), up to
order-one tensor normalisation factors. If \(N_{\rm KK}\) modes are
kinematically accessible, an inclusive rate receives the parametric
enhancement
\begin{equation}
  \sum_{n\leq N_{\rm KK}} g_n^2
  \sim
  \frac{N_{\rm KK}}{\bar{M}_{\rm Pl}^2},
  \label{eq:inclusive_coupling_framework}
\end{equation}
whereas a mode-resolved communication protocol remains governed by
the per-mode Planck suppression. For the flat interval,
\begin{equation}
  N_{\rm KK}(\omega)
  \sim
  \frac{\omega L}{\pi},
  \qquad
  \omega\gg \frac{\pi}{L}.
  \label{eq:NKK_framework}
\end{equation}
This distinction is central below: broadband incoherent sources may
benefit from the density of KK final states, while coherent addressing
of individual KK levels remains Planck-suppressed.

The sum in Eq.~\eqref{eq:GAB} must be understood as an
effective-field-theory expression. In a complete ultraviolet theory,
sufficiently massive KK modes are affected by the finite thickness of
the branes, string form factors, higher-derivative corrections,
warping, or strong-gravity effects near the fundamental scale. We
parametrise these effects by a dimensionless ultraviolet form factor
$F_n$ multiplying each overlap:
\begin{equation}
  f_n(y_A)\,f_n(y_B)
  \;\longrightarrow\;
  f_n(y_A)\,f_n(y_B)\,F_n .
  \label{eq:UV_form_factor}
\end{equation}
We assume
\begin{equation}
\begin{aligned}
  F_n &\simeq 1
  && \text{for } m_n\ll M_{\rm QG},
  \\
  F_n &\ll 1
  && \text{for } m_n\gg M_{\rm QG}.
\end{aligned}
  \label{eq:UV_form_factor_limits}
\end{equation}
The transition between these regimes need not be sharp. Here
$M_{\rm QG}$ denotes the ultraviolet scale at which the
higher-dimensional effective description ceases to be valid. 
In a string or M-theory compactification, $M_{\rm QG}$ should be
understood as the scale at which string, higher-dimensional Planck,
warping, or other ultraviolet effects invalidate the simple KK
effective description. We remain agnostic about its precise value. The factor
$F_n$ parametrises the breakdown of the effective KK description near
$M_{\rm QG}$; it is not specified further, because it is a property of
the ultraviolet completion and of the microscopic structure of the
branes. The only assumption used below is that the modes relevant for
the low-energy channel satisfy $m_n\ll M_{\rm QG}$, so that
$F_n\simeq1$.

With this caveat, the regulated transfer function is
\begin{equation}
  G_{AB}(p)
  =
  \sum_n
  f_n(y_A)\,f_n(y_B)\,F_n\,
  \frac{1}{p^2-m_n^2+i\epsilon}.
  \label{eq:GAB_regulated}
\end{equation}
Equation~\eqref{eq:GAB_regulated} contains the essential physics of
inter-brane communication. The signal transmitted from
$\mathcal{B}_A$ to $\mathcal{B}_B$ is controlled by the gravitational
coupling, the KK spectrum, the brane overlap factors, and the
ultraviolet form factor. A necessary condition for the channel to
exist is that at least part of the gravitational spectrum has
non-vanishing support on both branes---a condition satisfied by every
level of the flat-interval tower. In a general compactification this
becomes a question about the eigenfunctions of the internal Laplacian
and the positions of the branes.

From the perspective of communication theory,
Eq.~\eqref{eq:GAB_regulated} is the transfer function of the channel.
The compactification geometry determines the spectrum of available
communication modes; the overlap factors determine their coupling to
the two branes; and the ultraviolet form factor determines the range
over which the effective description is valid. The following sections
characterise the bandwidth, attenuation, mode structure, and
information-carrying capacity of this higher-dimensional gravitational
link.




\section{The Gravitational Channel}
\label{sec:channel}

The previous section identified the brane-to-brane propagator as the
basic transfer function of the inter-brane link. We now translate this
spectral object into the language of signal propagation. Throughout
this section we work in natural units \(c=\hbar=1\). When needed, SI
units are restored by writing
\begin{equation}
  k_n
  =
  \sqrt{
    \frac{\omega^2}{c^2}
    -
    \frac{m_n^2c^2}{\hbar^2}
  },
  \label{eq:kn_SI}
\end{equation}
when \(m_n\) is expressed as a mass.

The relevant question is not only which KK modes exist, but which of
them can carry a signal from a source on \(\mathcal{B}_A\) to a
detector on \(\mathcal{B}_B\) at a finite spatial separation. Consider
a source on \(\mathcal{B}_A\) oscillating at angular frequency
\(\omega\) and localised near \(\mathbf{x}_A\). A detector on
\(\mathcal{B}_B\) is located at \(\mathbf{x}_B\). We denote the
ordinary three-dimensional separation vector and its magnitude by
\begin{equation}
  \boldsymbol{\rho}
  =
  \mathbf{x}_B-\mathbf{x}_A,
  \qquad
  \rho = |\boldsymbol{\rho}| .
  \label{eq:rho_definition}
\end{equation}
At linear order, the induced response on \(\mathcal{B}_B\) is obtained
by convolving the source stress-energy tensor with the retarded
brane-to-brane Green function,
\begin{equation}
  h_{\mu\nu}^{(B)}(\omega,\mathbf{x}_B)
  =
  \int d^3\mathbf{x}'\,
  G^{R\,\mu\nu\rho\sigma}_{AB}
  (\omega;\mathbf{x}_B-\mathbf{x}')
  \,
  T^{(A)}_{\rho\sigma}(\omega,\mathbf{x}') .
  \label{eq:channel_response}
\end{equation}
As in the previous section, we suppress the tensor numerator whenever
only the spectral and propagation structure is relevant. The scalar
retarded kernel is the position-space representation of \(G_{AB}(p)\)
from Eq.~\eqref{eq:GAB_regulated}: its momentum-space form is obtained
by inverse Fourier transformation with respect to the three-momentum
\(\mathbf{k}\) at fixed \(\omega\). It may be written as a sum over KK
levels,
\begin{equation}
  G^R_{AB}(\omega,\boldsymbol{\rho})
  =
  \sum_n
  f_n(y_A)\,f_n(y_B)\,F_n\,
  G^R_n(\omega,\boldsymbol{\rho}) ,
  \label{eq:GAB_retarded_sum}
\end{equation}
where \(F_n\) is the ultraviolet form factor of
Eq.~\eqref{eq:UV_form_factor}. The four-dimensional retarded Green
function of the \(n\)-th KK mode satisfies the Helmholtz equation
\begin{equation}
  \left(\nabla^2+k_n^2\right)
  G^R_n(\omega,\boldsymbol{\rho})
  =
  -\delta^{(3)}(\boldsymbol{\rho}),
  \label{eq:Helmholtz_KK}
\end{equation}
with outgoing boundary conditions and wave number
\begin{equation}
  k_n(\omega)
  =
  \sqrt{\omega^2-m_n^2},
  \qquad
  \mathrm{Im}\,k_n\geq0.
  \label{eq:kn_definition}
\end{equation}
The outgoing solution depends only on
\(\rho=|\boldsymbol{\rho}|\) and is
\begin{equation}
  G^R_n(\omega,\rho)
  =
  \frac{e^{ik_n \rho}}{4\pi \rho}.
  \label{eq:KK_retarded_kernel}
\end{equation}

Equation~\eqref{eq:kn_definition} determines whether the \(n\)-th KK
level propagates or decays.

\textit{Propagating regime.} For \(\omega > m_n\), \(k_n\) is real and
the \(n\)-th level contributes an outgoing spherical wave,
\begin{equation}
  G^R_n(\omega,\rho)
  =
  \frac{e^{i\sqrt{\omega^2-m_n^2}\,\rho}}{4\pi \rho},
  \label{eq:propagating_mode}
\end{equation}
decaying only with the geometric \(1/\rho\) factor of
three-dimensional radiation. At threshold, \(k_n\to0\) and
\(v_{g,n}\to0\), so the radiative flux vanishes in the threshold
limit. A useful long-distance channel requires \(\omega\) to lie
sufficiently above the threshold that the group velocity is
non-negligible.

\textit{Evanescent regime.} For \(\omega < m_n\),
\(k_n=i\sqrt{m_n^2-\omega^2}\), and the contribution decays
exponentially,
\begin{equation}
  G^R_n(\omega,\rho)
  =
  \frac{e^{-\sqrt{m_n^2-\omega^2}\,\rho}}{4\pi \rho}.
  \label{eq:evanescent_mode}
\end{equation}
Such a mode can affect the near-field response at distances
\(\rho \lesssim1/\sqrt{m_n^2-\omega^2}\), but provides no long-range
radiative communication channel.

The number of radiative KK subchannels at frequency \(\omega\) is the
number of levels satisfying \(m_n<\omega\). For the flat interval,
\(m_n=n\pi/L\), so
\begin{equation}
  N_{\rm KK}(\omega)
  =
  \left\lfloor\frac{\omega L}{\pi}\right\rfloor.
  \label{eq:Nkk_channel}
\end{equation}
This is a kinematic count of open radiative modes; the number of usable
communication channels is generically smaller, because of finite
linewidths, detector resolution, cross-talk, noise, and the
ultraviolet form factors \(F_n\). The threshold frequencies at which
successive channels open are, in SI units,
\begin{equation}
  \nu_n
  =
  \frac{nc}{2L}.
  \label{eq:KK_frequency_thresholds_SI}
\end{equation}
For the benchmark \(L=0.1~\mu{\rm m}\),
\begin{equation}
  \nu_1
  \simeq
  1.5\times10^{15}~{\rm Hz},
  \qquad
  m_1c^2
  \simeq
  6~{\rm eV}.
  \label{eq:m1benchmark_channel}
\end{equation}
Below this threshold the only radiative channel is the massless
graviton. Above it, the tower opens level by level.

It is useful to relate this benchmark to a possible ultraviolet string
scale. Consider a weakly coupled string compactification with one large
compact dimension of length
\begin{equation}
  L=\pi R=0.1~\mu{\rm m},
\end{equation}
and with the remaining five internal radii of order the string length.
Using the standard volume relation for the four-dimensional Planck
mass, one obtains parametrically
\begin{equation}
  M_{\rm Pl}^2
  \sim
  \frac{4}{g_s^2}\,M_s^3 R ,
\end{equation}
where \(M_s\) is the string scale and \(g_s\) the string coupling.
Thus
\begin{equation}
  M_s
  \simeq
  \left(
    \frac{\pi g_s^2 M_{\rm Pl}^2}{4L}
  \right)^{1/3}
  \simeq
  6.1\times10^9~{\rm GeV}\,
  g_s^{2/3},
  \label{eq:string_scale_benchmark}
\end{equation}
where \(M_{\rm Pl}=1.22\times10^{19}~{\rm GeV}\) has been used. For
example, for \(g_s=0.1\),
\begin{equation}
  M_s
  \simeq
  1.3\times10^9~{\rm GeV}
  \simeq
  1.3\times10^6~{\rm TeV}.
\end{equation}
This estimate is sensitive to the sizes of the remaining compact
dimensions. If the other five radii are larger than the string length
by a common factor \(\xi\), the compact volume is enhanced by
\(\xi^5\), and the inferred string scale is reduced to
\begin{equation}
  M_s(\xi)
  \simeq
  6.1\times10^9~{\rm GeV}\,
  g_s^{2/3}\,\xi^{-5/3}.
  \label{eq:string_scale_benchmark_xi}
\end{equation}
For \(\xi=10\), this gives
\begin{equation}
  M_s
  \simeq
  1.3\times10^8~{\rm GeV}\,
  g_s^{2/3},
\end{equation}
or, for \(g_s=0.1\),
\begin{equation}
  M_s
  \simeq
  2.8\times10^7~{\rm GeV}
  \simeq
  2.8\times10^4~{\rm TeV}.
\end{equation}
Thus the benchmark compactification length corresponds to a very high
ultraviolet scale in simple string-motivated estimates, but the
precise number is not universal: it depends on the string coupling and
on the sizes of the remaining compact dimensions.

Since the flat three-dimensional part is translationally and
rotationally invariant, the scalar kernel depends only on
\(\rho=|\boldsymbol{\rho}|\). For the flat interval, with \(F_0\equiv1\) since
\(m_0=0\ll M_{\rm QG}\) in the effective theory, and with \(k_n\)
defined by Eq.~\eqref{eq:kn_definition}, the scalar retarded transfer
function can be written uniformly as
\begin{equation}
  G^R_{AB}(\omega,\rho)
  =
  \frac{1}{L}
  \frac{e^{i\omega \rho}}{4\pi \rho}
  +
  \frac{2}{L}
  \sum_{n\geq1}
  (-1)^n F_n\,
  \frac{e^{ik_n \rho}}{4\pi \rho},
  \label{eq:GAB_retarded_flat}
\end{equation}
where each \(k_n\) is real for propagating levels and purely imaginary
for evanescent ones. The alternating sign \((-1)^n\) is the relative
phase of the \(n\)-th wavefunction between the two endpoints. It plays
no role for a single isolated mode, but contributes to the phase
structure of the channel when several KK levels are excited
coherently.

Each propagating KK level has its own wave number \(k_n\) and group
velocity
\begin{equation}
  v_{g,n}
  =
  \frac{\partial\omega}{\partial k_n}
  =
  \frac{k_n}{\omega}
  =
  \sqrt{1-\frac{m_n^2}{\omega^2}}.
  \label{eq:group_velocity_KK}
\end{equation}
Massive KK modes close to threshold propagate slowly; modes with
\(\omega\gg m_n\) propagate nearly at \(c\). The channel is therefore
strongly dispersive near each threshold and weakly dispersive far
above it. In natural units, a pulsed source exciting several KK levels
simultaneously will produce arrivals on \(\mathcal{B}_B\) separated by
mode-dependent time delays
\begin{equation}
  \Delta t_n(\rho)
  \simeq
  \rho\!\left(\frac{1}{v_{g,n}}-1\right)
  \label{eq:KK_delay}
\end{equation}
relative to the massless zero mode. This dispersion is not merely a
nuisance: the arrival-time pattern is a signature of the KK mass
spectrum and could in principle help identify which modes were
populated.

The channel therefore has two complementary aspects. Spectrally, new
radiative subchannels open whenever the source frequency crosses a KK
threshold. Dynamically, each open subchannel carries a distinct phase,
group velocity, and attenuation, all fixed by the compactification
geometry through the masses \(m_n\), the overlap factors
\(f_n(y_A)f_n(y_B)\), and the form factors \(F_n\).

This is the sense in which the KK tower provides more than a set of
additional propagating modes. It enlarges the signal space. A
transmitter need not encode information only in the amplitude or phase
of a single gravitational waveform: it may encode information in the
distribution of excitation across the accessible KK levels, in their
relative phases, or in the resulting mode-dependent arrival pattern on
\(\mathcal{B}_B\). The feasibility of any such encoding depends on the
source and detector physics; the linear propagation kernel is already
determined by the retarded Green function
Eq.~\eqref{eq:GAB_retarded_flat}.



\section{Power Transmission Through the Bulk}
\label{sec:power}

The existence of a propagating KK mode does not by itself guarantee
an efficient communication channel. The previous section described the
linear propagation kernel between the two branes. We now ask a more
operational question: how much of the gravitational radiation produced
on \(\mathcal{B}_A\) can be delivered to, and in principle recovered by,
observers on \(\mathcal{B}_B\)?

At the linearised level, the brane-to-brane Green function transfers
field amplitudes. Power is obtained only after specifying the source,
the radiative flux, the detector response, and the bandwidth over
which the signal is measured. We therefore do not assign a universal
transmission probability to the bulk. Instead, we describe the
parametric structure of the power transfer.

Let a source on \(\mathcal{B}_A\) emit gravitational radiation with
spectral power distribution \(dP_A/d\omega\). The fraction of this
radiation that can reach \(\mathcal{B}_B\) through the bulk decomposes
into KK contributions,
\begin{equation}
  \frac{dP_{\rm bulk}}{d\omega}
  =
  \frac{dP_0}{d\omega}
  +
  \sum_{n\geq1}
  \frac{dP_n}{d\omega},
  \label{eq:bulk_power_decomp}
\end{equation}
where \(P_0\) denotes emission into the massless graviton and \(P_n\)
emission into the \(n\)-th KK graviton. The branching fractions are
source-dependent: they depend on the stress-energy tensor of the
emitter, the tensor polarisation structure, the KK wavefunction at the
emitting brane, possible greybody factors, and the ultraviolet form
factor \(F_n\). They are not determined by kinematics alone.

For a detector on \(\mathcal{B}_B\) at ordinary three-dimensional
distance \(\rho\), the amplitude associated with the \(n\)-th KK mode
is proportional to
\begin{equation}
  \mathcal{A}_n(\omega,\rho)
  \;\propto\;
  f_n(y_A)\,f_n(y_B)\,F_n\,
  \frac{e^{ik_n \rho}}{4\pi \rho},
  \label{eq:mode_amplitude_power}
\end{equation}
with \(k_n\) defined in Eq.~\eqref{eq:kn_definition}. For propagating
modes \(k_n\) is real and the radiative flux falls geometrically as
\(1/\rho^2\). For evanescent modes \(k_n\) is purely imaginary and the
contribution is exponentially suppressed. The relevant set of
long-distance power-carrying modes is therefore the set of open
levels, \(m_n<\omega\).

A schematic expression for the received spectral power is
\begin{equation}
  \frac{dP_B}{d\omega}
  =
  \sum_{n,m}
  \mathcal{R}_{nm}(\omega)\,
  \mathcal{A}_n(\omega,\rho)\,
  \mathcal{A}_m^\ast(\omega,\rho)\,
  S_{nm}(\omega),
  \label{eq:received_power_general}
\end{equation}
where \(\mathcal{R}_{nm}\) is the receiver response matrix, encoding
the tensor projection, frequency response, and angular acceptance of
the detector. The matrix \(S_{nm}(\omega)\) is the cross-spectral
density of the source in KK space. Its diagonal entries are the
spectral powers emitted into the individual KK modes,
\begin{equation}
  S_{nn}(\omega)
  =
  \frac{dP_n}{d\omega},
  \label{eq:source_spectral_diagonal}
\end{equation}
while its off-diagonal entries encode correlations between different
KK amplitudes. For statistically independent emissions into distinct
KK levels,
\begin{equation}
  S_{nm}(\omega)
  =
  \delta_{nm}\,
  \frac{dP_n}{d\omega},
  \label{eq:source_spectral_incoherent}
\end{equation}
and the received power reduces to an incoherent sum over modes. For a
fully coherent source, the off-diagonal entries may instead be of
order
\begin{equation}
  S_{nm}(\omega)
  \sim
  e^{i\phi_{nm}(\omega)}
  \sqrt{
    \frac{dP_n}{d\omega}
    \frac{dP_m}{d\omega}
  },
  \qquad n\neq m,
  \label{eq:source_spectral_coherent}
\end{equation}
with phases \(\phi_{nm}\) determined by the source preparation. Thus
Eq.~\eqref{eq:received_power_general} does not assume either coherent
or incoherent emission; it displays the general structure in which
source correlations, propagation phases, and detector projections all
enter separately.

Equation~\eqref{eq:received_power_general} is not a universal detector
formula. Its purpose is to make explicit the roles of emission,
propagation, and detection before imposing any specific source or
receiver model.

Two limiting cases are useful for the later information-theoretic
discussion.

\paragraph{Resolved incoherent modes.}
If the receiver can resolve neighbouring KK levels spectrally, and if
the source or measurement averages over relative phases between modes,
the off-diagonal terms in Eq.~\eqref{eq:received_power_general} are
suppressed. The received power reduces schematically to
\begin{equation}
  \frac{dP_B}{d\omega}
  \simeq
  \sum_{n\,:\,m_n<\omega}
  \eta_n(\omega,\rho)\,
  \frac{dP_n}{d\omega},
  \label{eq:received_power_incoherent}
\end{equation}
where
\begin{equation}
  \eta_n(\omega,\rho)
  \sim
  |f_n(y_A)\,f_n(y_B)|^2\,
  |F_n|^2\,
  \frac{\mathcal{R}_{nn}(\omega)}{(4\pi \rho)^2},
  \label{eq:eta_n}
\end{equation}
up to tensor and normalisation factors. For the flat interval,
\(|f_n(y_A)f_n(y_B)|^2 = 4/L^2\) for all \(n\geq1\), so the
mode-dependence of \(\eta_n\) is carried entirely by \(|F_n|^2\) and
\(\mathcal{R}_{nn}\). In this regime the KK tower acts as a set of
approximately independent power channels, which is the picture
employed in the information-theoretic discussion of
Section~\ref{sec:capacity}.

\paragraph{Coherent multimode response.}
If several KK modes are excited with fixed relative phases and the
receiver measures them coherently, the off-diagonal terms in
Eq.~\eqref{eq:received_power_general} cannot be neglected. The
received power then contains interference contributions,
\begin{equation}
\begin{aligned}
  \mathcal{A}_n\mathcal{A}_m^\ast
  &\propto
  \bigl[f_n(y_A)f_n(y_B)\bigr]\,
  \bigl[f_m(y_A)f_m(y_B)\bigr]\,
  F_n F_m^\ast
  \\
  &\hspace{1.5cm}\times
  e^{i(k_n-k_m)\rho},
  \qquad n\neq m .
\end{aligned}
  \label{eq:mode_interference}
\end{equation}
For the flat interval, the endpoint overlaps carry alternating signs
\((-1)^n\), contributing to this coherent phase structure. Such
interference can enhance or suppress the response at a given
frequency and distance, and is sensitive to source coherence,
propagation phases, and detector resolution. This is the regime in
which the KK tower behaves as a genuine multi-mode linear channel
rather than a simple sum of independent power contributions.

The distinction matters physically. A stochastic source, such as
a thermal emitter, naturally leads to an incoherent sum over
accessible modes. A phase-controlled source, such as coherent
photon--graviton conversion, can in principle prepare a structured
multimode state. These two cases lead to different communication
strategies and to different capacity bounds.

Three frequency regimes are useful for orientation.

\paragraph{Low-frequency regime.}
For \(\omega\ll m_1=\pi/L\), all massive KK modes are evanescent at
macroscopic distance. The only radiative channel is the massless
graviton, and the inter-brane link reduces to ordinary
four-dimensional gravitational radiation. Communication in this
regime is zero-mode communication, regardless of the source
luminosity.

\paragraph{Threshold regime.}
When \(\omega\) exceeds a KK threshold \(m_n\), a new radiative channel
opens. However, as shown in Section~\ref{sec:channel}, the group
velocity vanishes at threshold and the transported radiative flux is
negligible there. More quantitatively, requiring a minimum group
velocity \(v_{g,n}>v_\ast\) gives
\begin{equation}
  \omega >
  \frac{m_n}{\sqrt{1-v_\ast^2}} .
\end{equation}
For example, \(v_{g,n}>0.5\) requires \(\omega\gtrsim1.15\,m_n\),
whereas \(v_{g,n}>0.9\) requires \(\omega\gtrsim2.3\,m_n\).

In the ideal flat bulk, the KK modes are stable propagation
eigenstates. Bulk interactions obey KK momentum selection rules, and
for an exactly linear spectrum \(m_n=n\pi/L\) the decay of one KK mode
into lower KK modes is either forbidden by KK momentum conservation or
pushed to zero phase space. Finite widths therefore require effects
that violate the ideal KK momentum selection rule or modify the
spectrum: brane-localised interactions, brane-localised loops,
localised kinetic terms, warping, curvature, finite brane thickness,
or other ultraviolet effects. When such effects are present, the
\(n\)-th mode may acquire an effective width \(\Gamma_n^{\rm eff}\).

A mode-resolved description is meaningful only when neighbouring
resonances remain spectrally distinguishable,
\begin{equation}
  \Gamma_n^{\rm eff} \ll \Delta m_n,
  \label{eq:resolved_width_condition}
\end{equation}
where, for the flat interval, the level spacing
\begin{equation}
  \Delta m_n = m_{n+1}-m_n = \frac{\pi}{L}
\end{equation}
is mode-independent. In the strict idealised limit
\(\Gamma_n^{\rm eff}\to 0\), the KK levels are infinitely sharp
propagation channels. In a realistic brane compactification, small but
nonzero widths may be induced by localised or ultraviolet physics. We
leave their calculation model-dependent.

\paragraph{Multimode regime.}
For \(\omega\gg\pi/L\), many KK levels are kinematically open, with
\begin{equation}
  N_{\rm KK}(\omega)
  \sim
  \frac{\omega L}{\pi}
  \label{eq:Nkk_power}
\end{equation}
up to the ultraviolet cutoff encoded by the form factors \(F_n\). Not
all open levels are equally useful: some may couple weakly to the
source, some may be poorly resolved by the detector, and modes near
the ultraviolet scale may be suppressed by the microscopic brane
structure. The robust qualitative point is that increasing the source
frequency opens a larger signal space.

\paragraph{Numerical benchmark.}
The first KK threshold for \(L=0.1~\mu{\rm m}\) is, as in
Eq.~\eqref{eq:m1benchmark_channel},
\begin{equation}
  \nu_1
  \simeq
  1.5\times10^{15}~{\rm Hz},
  \qquad
  m_1c^2
  \simeq
  6~{\rm eV}.
  \label{eq:m1benchmark_power}
\end{equation}
By contrast, compact-object gravitational-wave sources have
characteristic frequencies \(\nu_{\rm GW}\sim10^2\)--\(10^4~{\rm Hz}\),
giving
\begin{equation}
  \frac{\nu_1}{\nu_{\rm GW}}
  \sim
  10^{11}\text{--}10^{13}.
  \label{eq:threshold_ratio_power}
\end{equation}
Conventional astrophysical gravitational-wave sources are therefore
powerful zero-mode emitters that do not access the KK tower at this
benchmark scale. This gap is also many orders of magnitude above the
range of current and proposed high-frequency gravitational-wave
searches, which focus on MHz--GHz frequencies and already face
substantial experimental challenges~\cite{Aggarwal:2020olq}.
Accessing the KK tower therefore requires sources qualitatively
different from those encountered in ordinary gravitational-wave
astronomy.

A useful transmitter must satisfy two separate conditions: it must
couple power into gravitational degrees of freedom, and it must do so
at frequencies high enough to populate KK modes. The following section
discusses several candidate source classes from this standpoint.


\section{Candidate Transmitters}
\label{sec:emitters}

The preceding sections characterised the bulk gravitational channel
independently of any specific source. We now ask how an emitter on
$\mathcal{B}_A$ might populate that channel. The purpose of this
section is not to propose an engineering design. It is to classify
physical source mechanisms according to the kind of gravitational
input state they could, in principle, prepare.

A useful transmitter must satisfy two distinct requirements. First, it
must couple energy into gravitational degrees of freedom. Second, it
must do so at frequencies high enough to access the relevant part of
the KK spectrum. These two requirements are logically independent:
large gravitational luminosity at low frequency excites only the
massless zero mode, while a high-frequency source with negligible
gravitational branching may access the tower only extremely weakly.

We organise candidate transmitters by three criteria:
\begin{enumerate}
\item whether the characteristic frequency lies below or above the
first KK threshold;
\item whether the source excites one mode, a few resolved modes, or a
broad range of KK levels;
\item whether the emitted state is coherent and controllable, or
thermal and stochastic.
\end{enumerate}
For the benchmark value $L=0.1~\mu{\rm m}$, the first KK threshold is
(cf.\ Eq.~\eqref{eq:m1benchmark_channel})
\begin{equation}
  \nu_1 \simeq 1.5\times10^{15}~{\rm Hz},
  \qquad
  m_1c^2\simeq6~{\rm eV}.
  \label{eq:m1emitters}
\end{equation}
This is only a reference scale; for a different compactification
length all frequencies rescale as $L^{-1}$.

\subsection{Compact-object mergers: high power, zero-mode dominated}

Compact binaries are the most powerful gravitational-wave sources
known. Their characteristic frequency near merger is, up to
numerical factors,
\begin{equation}
  \nu_{\rm GW} \sim \frac{c^3}{GM}.
  \label{eq:nugwcompact}
\end{equation}
For stellar-mass black holes, $M\sim5$--$100\,M_\odot$, this gives
$\nu_{\rm GW}\sim10^2$--$10^4~{\rm Hz}$; neutron-star mergers reach
at most a few times $10^4~{\rm Hz}$. These frequencies lie many
orders of magnitude below the first KK threshold of
Eq.~\eqref{eq:m1emitters}.

Compact mergers therefore illustrate the high-power, low-frequency
limit. They radiate enormous power into the massless graviton, but
do not kinematically access the KK tower at the benchmark
compactification scale. Their radiation is effectively
four-dimensional. This example separates gravitational luminosity
from KK bandwidth: power alone is not sufficient for inter-brane
communication through massive modes.

\subsection{Black holes as thermal broadband emitters}

Black holes provide a qualitatively different class of sources
because their emission is controlled by a temperature rather than
by an orbital frequency. We consider black holes localised on
$\mathcal{B}_A$. The four-dimensional formulae below are used only
to identify scales; when the horizon radius approaches $L$, the
higher-dimensional geometry and greybody factors become
model-dependent.

The Hawking temperature is
\begin{equation}
  T_H = \frac{\hbar c^3}{8\pi G M k_B}.
  \label{eq:HawkingT_emitters}
\end{equation}
Thermal emission into the $n$-th KK level is Boltzmann suppressed
unless
\begin{equation}
  k_B T_H \gtrsim m_n c^2.
  \label{eq:Hawking_KK_condition}
\end{equation}
This is a necessary kinematic condition; the actual emission rate
also depends on greybody factors whose computation requires the
full higher-dimensional geometry. For the first KK level of the
benchmark, $m_1c^2\simeq6~{\rm eV}$, the condition
Eq.~\eqref{eq:Hawking_KK_condition} corresponds to
$T_H\gtrsim7\times10^4~{\rm K}$, or
\begin{equation}
  M
  \lesssim
  2\times10^{18}~{\rm kg}
  \left(\frac{L}{0.1~\mu{\rm m}}\right).
  \label{eq:MhawkingKK_emitters}
\end{equation}
The scaling $M_{\rm Hawking}\propto L$ reflects the fact that a
larger $L$ lowers $m_1$, making the threshold easier to reach with
a cooler (and hence more massive) black hole. When
$k_BT_H\gg m_1c^2$, the number of thermally accessible KK levels
is of order
\begin{equation}
  N_{\rm KK}^{(H)} \sim \frac{k_BT_H}{m_1c^2}.
  \label{eq:NkkH_emitters}
\end{equation}

Only the bulk gravitational component of the Hawking radiation can
propagate to $\mathcal{B}_B$. Most emitted quanta may remain
confined to $\mathcal{B}_A$ if they are brane-localised species.
The branching fraction into bulk modes depends on the number of
brane and bulk degrees of freedom, greybody factors, the horizon
geometry, and the compactification. We do not compute it here.

A low-mass black hole is therefore a broadband but incoherent
transmitter. It can populate many KK modes simultaneously, but
the emission is thermal and carries no phase control. Information
could only be encoded in macroscopic parameters of the source,
not in a coherent waveform.

\begin{table*}[t]
\centering
\setlength{\tabcolsep}{8pt}
\renewcommand{\arraystretch}{1.3}
\begin{tabular}{l|c|c|c}
\hline\hline
Source class
& Power into bulk
& KK-sector access
& Signal control
\\
\hline
Compact mergers
& Very high zero-mode power
& None for benchmark $L$
& None
\\
Low-mass black holes
& Model-dependent
& Broad, thermal
& Poor
\\
Artificial BH bursts
& High burst power in principle
& Broad, thermal
& Timing and coarse spectrum
\\
Photon--graviton conversion
& Negligible per mode
& Tunable, coherent
& High
\\
\hline\hline
\end{tabular}
\caption{Qualitative comparison of candidate transmitter regimes
for the benchmark compactification $L=0.1~\mu{\rm m}$.
Compact mergers illustrate high gravitational luminosity at
frequencies far below the first KK threshold. Black-hole sources
illustrate broadband thermal emission. Photon--graviton conversion
illustrates coherent but extremely weak mode-selective excitation.
All entries are qualitative; none constitutes an engineering
proposal.}
\label{tab:emitters}
\end{table*}

\subsection{Artificial black-hole bursts: extreme energy concentration}

One may consider a further idealisation in which short-lived black
holes are produced with controlled macroscopic parameters. Such
objects should not be tied to one specific production mechanism.
The essential requirement is gravitational compactness: an energy
$E$ must be confined inside a region smaller than the corresponding
Schwarzschild radius,
\begin{equation}
  R \lesssim R_s(E) = \frac{2GE}{c^4} = \frac{2GM}{c^2},
  \label{eq:compactness_BH}
\end{equation}
where $M=E/c^2$. In a higher-dimensional compactification the
trapped-surface criterion is modified, but the physical requirement
is the same: energy must be compressed into a sufficiently small
region for gravitational collapse to occur.

This formulation is more general than the collider picture. A
high-energy collision is one idealisation, especially for producing
the smallest possible black holes, in which case the relevant scale
is the ultraviolet quantum-gravity scale $M_{\rm QG}$. A black hole
can also be semiclassical, with radius much larger than the
microscopic quantum-gravity length, in which case its formation is
governed by macroscopic compactness rather than a
centre-of-mass-energy threshold. We do not assume any specific
formation mechanism.

For an artificial source, one could imagine the compression of
matter or radiation into a sufficiently small volume. We do not
claim that such a process is realistic. It would face severe
obstacles, including diffraction, pressure support, opacity, plasma
formation, pair production, hydrodynamic instabilities, and
gravitational backreaction. The point is only to define a limiting
class of burst-like sources whose input alphabet could be
represented schematically as
\begin{equation}
  \{M_i,J_i,Q_i,t_i\},
  \label{eq:BHcodeparams}
\end{equation}
where $M_i$, $J_i$, $Q_i$, and $t_i$ denote the mass, angular
momentum, charge, and creation time of the $i$-th object. These
parameters determine the Hawking temperature and the number of
thermally accessible KK modes:
\begin{equation}
  M_i \;\longrightarrow\; T_H(M_i) \;\longrightarrow\;
  N_{\rm KK}^{(H)}(M_i).
  \label{eq:BHthermalencoding}
\end{equation}
This is not a coherent antenna. It is a stochastic burst transmitter:
the message is encoded in the macroscopic preparation, while the
channel output is a thermal distribution of brane and bulk quanta.
Its role here is conceptual: it illustrates a source that is
broadband in KK space but poor in phase control.

\subsection{Photon--graviton conversion: coherent but extremely weak}

The most controllable class of candidate transmitters is
electromagnetic. In the Gertsenshtein effect, an electromagnetic
wave in a static magnetic field converts into a graviton,
\begin{equation}
  \gamma + B_{\rm ext} \longrightarrow g,
  \label{eq:Gertsenshtein}
\end{equation}
and, in the higher-dimensional setting, into a massive KK graviton,
\begin{equation}
  \gamma + B_{\rm ext} \longrightarrow g_n .
  \label{eq:KKGertsenshtein}
\end{equation}
The process is superficially analogous to photon--axion conversion,
but the coupling is gravitational: it is to the energy-momentum
tensor and is suppressed by the four-dimensional Planck scale. The
conversion probability is therefore extraordinarily small.

For an interaction length $\ell$ in an external magnetic field $B$, the
conversion probability to the $n$-th KK mode scales parametrically as
\begin{equation}
  P_{\gamma\to g_n}
  \sim
  \frac{\hbar}{\mu_0 c^3}
  \left(\frac{B\ell}{\bar M_{\rm Pl}}\right)^2
  \left|
  \frac{\sin(\Delta k\,\ell/2)}
       {\Delta k\,\ell/2}
  \right|^2
  \mathcal{O}_n ,
  \label{eq:Pgrav}
\end{equation}
where \(B\) is expressed in tesla, \(\ell\) in metres,
\(\bar M_{\rm Pl}\) in kilograms, and \(\mathcal{O}_n\) is a
dimensionless overlap and polarisation factor. In natural
Heaviside--Lorentz units, this reduces to the familiar scaling
\((B\ell/\bar M_{\rm Pl})^2\), up to convention-dependent order-one
polarisation factors. The physical phase mismatch is
\begin{equation}
  \Delta k
  =
  \frac{\omega}{c}
  -
  \sqrt{
    \frac{\omega^2}{c^2}
    -
    \frac{m_n^2c^2}{\hbar^2}
  } .
  \label{eq:Deltakgrav}
\end{equation}
In the ultrarelativistic regime \(\hbar\omega\gg m_nc^2\),
\begin{equation}
  \Delta k
  \simeq
  \frac{m_n^2c^3}{2\hbar^2\omega}.
  \label{eq:Deltakgrav_UR}
\end{equation}

Coherent conversion over an interaction length $\ell$ requires
$\Delta k\,\ell\ll1$, which is easier to satisfy far above threshold.
Near threshold, $\hbar\omega\simeq m_nc^2$, the KK graviton has small
group velocity, the phase mismatch is large on macroscopic baselines,
and the conversion is strongly suppressed. Efficient conversion to a
given KK level therefore requires operation sufficiently above
threshold, or an additional phase-matching mechanism.

For the benchmark $m_1c^2\simeq 6~{\rm eV}$, the first KK threshold
lies at ultraviolet energies, corresponding to a wavelength
$\lambda\simeq 200~{\rm nm}$. The decisive advantage of this class is
not power, but control: amplitude, frequency, phase, polarisation,
and time structure can all be modulated with high precision.

Photon--graviton conversion is the opposite extreme from black-hole
evaporation: coherent, spectrally selective, and in principle
compatible with mode-resolved encoding, but with negligibly small
per-mode power.

\subsection{Comparison of transmitter regimes}

The source classes above are summarised in Table~\ref{tab:emitters}.
They should be read as limiting regimes, not as engineering proposals.

The trade-off is clear. Astrophysical compact objects provide
enormous gravitational power, but at frequencies too low to access
the KK tower at the benchmark scale. Black-hole evaporation can
populate many KK levels, but incoherently and with model-dependent
branching into the bulk. Coherent electromagnetic conversion offers
fine control over the emitted state, but with an extremely small
gravitational conversion probability.

The rest of the paper does not rely on the practical feasibility of
any of these transmitters. We assume only that some source on
$\mathcal{B}_A$ prepares an input state in the bulk gravitational
channel. The following sections ask how such an input could be
encoded, propagated, and decoded.


\section{From Physical Propagation to Channel Variables}
\label{sec:signal}

The preceding sections described three ingredients of the inter-brane
link: the KK spectrum and overlap factors, the retarded propagation
kernel, and possible source mechanisms. We now translate these
ingredients into the variables used in the information-theoretic
description. The purpose of this section is not to rederive the
propagator, nor to model a specific detector. It is to identify which
parts of the physical response become channel gains, input variables,
and receiver outputs.

Throughout this section we work in natural units $c=\hbar=1$.

\subsection*{Linear response and detector projection}

At linear order, a stress-energy tensor source on \(\mathcal{B}_A\)
induces a metric perturbation on \(\mathcal{B}_B\) according to
\begin{equation}
  h_{\mu\nu}^{(B)}(\omega,\mathbf{x}_B)
  =
  \int d^3\mathbf{x}'\,
  G^{R\,\mu\nu\rho\sigma}_{AB}
  (\omega;\mathbf{x}_B-\mathbf{x}')
  \,
  T^{(A)}_{\rho\sigma}(\omega,\mathbf{x}') .
  \label{eq:hresponse}
\end{equation}
This is the same retarded response relation as
Eq.~\eqref{eq:channel_response}. The object \(G^R_{AB}\) is the
brane-to-brane Green function derived in Section~\ref{sec:channel}.
It contains the KK masses, endpoint overlaps, propagation phases,
evanescent suppression factors, and the ultraviolet form factors
\(F_n\).

Strictly speaking, \(h_{\mu\nu}\) is not a gauge-invariant local
observable. Equation~\eqref{eq:hresponse} should therefore be
understood in a fixed linearised gauge, or as a convenient proxy for
the response of a specified detector to gauge-invariant tidal fields.
The pole positions, KK thresholds, and mode-counting statements are
gauge-independent; the tensor numerator and the detector projection
are not. In the information-theoretic model below, all such projection
effects are absorbed into the channel matrix.

Let the transmitter control a finite-dimensional subspace of source
stress-energy tensors, spanned by waveforms
\(\psi_\alpha^{\rho\sigma}(\omega,\mathbf{x})\), and let the receiver
measure a finite set of detector observables \(D_a\). The controllable
part of the source stress-energy tensor is expanded as
\begin{equation}
  T_A^{\rho\sigma}(\omega,\mathbf{x})
  =
  \sum_\alpha
  s_\alpha(\omega)\,
  \psi_\alpha^{\rho\sigma}(\omega,\mathbf{x}) ,
  \label{eq:source_basis_expansion}
\end{equation}
so that the complex coefficients \(s_\alpha\) are the transmitted
signal amplitudes in the chosen source basis. Similarly, the receiver
outputs \(r_a\) are the projections of the induced metric response
\(h_{\mu\nu}^{(B)}\), or equivalently of the associated detector
response, onto the detector observables \(D_a\). Projecting
Eq.~\eqref{eq:hresponse} onto these bases gives
\begin{equation}
  r_a(\omega)
  =
  \sum_\alpha
  H_{a\alpha}(\omega)\,
  s_\alpha(\omega)
  +
  n_a(\omega),
  \label{eq:projected_response}
\end{equation}
where \(n_a\) denotes detector noise.

The channel matrix \(H_{a\alpha}\) is the corresponding
finite-dimensional projection of the physical Green function,
\begin{equation}
  H_{a\alpha}(\omega)
  =
  \left\langle
    D_a,\,
    G^R_{AB}(\omega)\,\psi_\alpha
  \right\rangle ,
  \label{eq:H_projection}
\end{equation}
where tensor indices on \(\psi_\alpha^{\rho\sigma}\) and
\(G^R_{AB}\) are suppressed, and the brackets denote the appropriate
spacetime, tensor, and detector-response contractions. This expression
is schematic, but it makes the hierarchy clear: the physical
propagator determines the linear kernel, while the choice of
transmitter and receiver determines the finite-dimensional channel.

\subsection*{KK-resolved channel variables}

When the receiver can resolve the KK spectrum, it is useful to choose
a transmission basis labelled by KK levels. A source operating above
the first threshold may then prepare amplitudes
\begin{equation}
  \mathbf{s}
  =
  (s_0,s_1,\ldots,s_{N_{\rm KK}}),
  \label{eq:KK_amplitude_vector}
\end{equation}
where \(s_n\) is the complex transmitted amplitude associated with
the \(n\)-th resolved mode. In the ideal flat interval, and for a
detector that projects onto the same basis, the leading scalar part of
the mode-resolved gain is
\begin{equation}
  H_n(\omega,\rho)
  \propto
  f_n(y_A)f_n(y_B)\,
  F_n\,
  \frac{e^{ik_n \rho}}{4\pi \rho},
  \label{eq:Hn_physical}
\end{equation}
where \(\rho\) is the ordinary three-dimensional propagation distance
along the brane, and \(k_n\) is defined in
Eq.~\eqref{eq:kn_definition}. Tensor polarisations, detector response,
finite bandwidth, and normalisation factors multiply this expression.

Equation~\eqref{eq:Hn_physical} should not be read as saying that all
open KK levels are equally useful. For the flat interval the endpoint
overlap \(|f_n(y_A)f_n(y_B)|\) is mode-independent for \(n\geq1\), but
the remaining factors are not. Modes close to threshold have small
group velocity, evanescent modes are exponentially suppressed, modes
near the ultraviolet scale can be reduced by \(F_n\), and any real
detector has a frequency-dependent response. These effects determine
which open modes become usable communication degrees of freedom.

In a general compactification the KK eigenfunctions need not be
simultaneously aligned with the transmitter and receiver bases. The
effective channel matrix may then be non-diagonal,
\begin{equation}
  r_m(\omega)
  =
  \sum_\ell H_{m\ell}(\omega)\,s_\ell(\omega)
  +
  n_m(\omega),
  \label{eq:KK_channel_general}
\end{equation}
where \(r_m\) is the received output in the \(m\)-th detector channel
and \(n_m\) denotes noise. The optimal communication modes are then
linear combinations of KK states rather than individual KK levels.

\subsection*{Coherent and incoherent use of the tower}

The same physical channel can be used in different statistical
regimes. If the source prepares fixed relative phases between modes
and the receiver preserves phase information, the off-diagonal
structure of \(H_{mn}\) matters. This is the coherent multimode regime
relevant for controlled sources such as photon--graviton conversion.

If instead the source is stochastic, as in thermal black-hole
emission, or if the measurement averages over relative phases, the
off-diagonal correlations are suppressed. The tower then behaves
approximately as a set of independent power channels, with rates
controlled by the diagonal gains \(|H_n|^2\) and by the noise in each
resolved frequency bin.

This distinction is central for the following sections. The KK tower
does not automatically provide independent communication channels.
Independence is an additional approximation requiring spectral
resolution, negligible cross-talk, and a detector basis in which the
noise covariance is sufficiently close to diagonal.

\subsection*{Geometry encoded in the channel}

The channel matrix contains geometric information about the compact
space. The thresholds determine the KK masses \(m_n\), while the
overlaps determine how strongly the corresponding eigenfunctions
connect the two branes. For the flat interval with endpoint branes,
the spacing
\begin{equation}
  m_{n+1}-m_n = \frac{\pi}{L}
\end{equation}
directly determines \(L\). In more general geometries, the measured
data take the form
\begin{equation}
  \{m_n,\ f_n(Y_A)f_n(Y_B),\ H_{mn}\},
  \label{eq:geometric_channel_data}
\end{equation}
where \(Y_A\) and \(Y_B\) denote the locations of the two branes in
the compact space. These data constrain both the compactification
geometry and the brane locations.

This is an inverse spectral problem, related in spirit to Kac's
question of whether one can hear the shape of a drum~\cite{Kac:1966}.
It is not, in general, uniquely solvable. Distinct compact
Riemannian manifolds can share the same Laplacian spectrum without
being isometric~\cite{Milnor:1964}. The channel data nevertheless
carry non-trivial information: they determine the accessible masses,
the brane-to-brane overlaps, and the effective communication basis.
Thus the compact geometry is not merely a background parameter. It is
part of the channel itself.

\section{Information-Theoretic Channel Model}
\label{sec:capacity}

We now formulate the inter-brane link as an information-theoretic
channel. The goal is not to compute a numerical capacity. That would
require a concrete source, detector, noise spectrum, bandwidth, and
power constraint. Instead, we identify the linear channel structure
and the assumptions under which the KK tower can be treated as a set
of communication resources.

\subsection*{Abstract linear channel}

After projection onto a finite-dimensional set of transmitter and
receiver modes, the physical response
Eq.~\eqref{eq:projected_response} takes the standard linear form
\begin{equation}
  \mathbf{r}(\omega)
  =
  H(\omega)\,\mathbf{s}(\omega)
  +
  \mathbf{n}(\omega).
  \label{eq:abstractchannel}
\end{equation}
Here \(\mathbf{s}\in\mathbb{C}^{N_{\rm tx}}\) is the vector of
transmitted signal amplitudes, \(\mathbf{r}\in\mathbb{C}^{N_{\rm rx}}\)
is the vector of receiver outputs,
\(H\in\mathbb{C}^{N_{\rm rx}\times N_{\rm tx}}\) is the projected
channel matrix, and \(\mathbf{n}\) is the receiver noise. The entries
of \(H\) are not postulated independently: they are obtained by
projecting the retarded Green function of Section~\ref{sec:channel}
onto the chosen source and detector bases.

The noise covariance is
\begin{equation}
  \Sigma(\omega)
  =
  \left\langle
    \mathbf{n}(\omega)\mathbf{n}^\dagger(\omega)
  \right\rangle .
  \label{eq:noisecov}
\end{equation}
In the absence of a detector model, \(\Sigma\) is an external input.
A capacity estimate requires specifying \(H\), the allowed input
covariance, the bandwidth, and \(\Sigma\) consistently.

\subsection*{Gaussian capacity bound}

For a Gaussian linear channel with input covariance
\(Q(\omega)=\langle\mathbf{s}\mathbf{s}^\dagger\rangle\geq0\) and
power constraint \(\mathrm{Tr}\,Q\leq P_{\rm tot}\), the maximum
mutual information per channel use at frequency \(\omega\) is
\begin{equation}
  C_{\rm use}(\omega)
  =
  \max_{\substack{Q\geq0\\ \mathrm{Tr}\,Q\leq P_{\rm tot}}}
  \log_2\det
  \!\left[
    I+\Sigma^{-1/2}H Q H^\dagger\Sigma^{-1/2}
  \right],
  \label{eq:MIMOcapacity_general}
\end{equation}
under the usual assumptions of Gaussian signalling and known channel
matrix~\cite{Telatar:1999}. With the standard communication-theory
normalisation, a frequency band of width \(B\) contributes a factor
\(B\) to the information rate; a scalar Gaussian channel gives
\(B\log_2(1+\mathrm{SNR})\) bits per second~\cite{Shannon}. In the
continuum limit, the total rate is written schematically as
\begin{equation}
  C_{\rm total}
  =
  \int_{\mathcal{B}}
  C_{\rm use}(\nu)\,d\nu,
  \label{eq:Ctotal}
\end{equation}
where \(\nu=\omega/(2\pi)\) and \(\mathcal{B}\) is the accessible
frequency band.

Equation~\eqref{eq:MIMOcapacity_general} is an idealised upper bound
for the inter-brane problem. It assumes a linear stationary channel,
Gaussian inputs, perfect knowledge of \(H\) at the transmitter and
receiver, and a specified noise covariance. None of these assumptions
is automatic for a gravitational detector.

\subsection*{Resolved KK limit}

Multiple propagating KK modes do not automatically constitute
independent communication channels. The resolved KK limit is the
regime in which four conditions hold simultaneously:
\begin{enumerate}
\item neighbouring KK levels are spectrally distinguishable;
\item mode mixing and cross-talk are negligible during the
communication time;
\item the receiver can project onto the relevant KK basis;
\item the noise covariance is approximately diagonal in that basis.
\end{enumerate}
These conditions are plausible in the ideal flat interval at moderate
frequencies, but they must be checked in any concrete compactification
and detector model.

In this limit the channel matrix is approximately diagonal,
\(H_{mn}\simeq\delta_{mn}H_n\), and the MIMO bound reduces to a
parallel-channel estimate,
\begin{equation}
  C_{\rm KK}
  \leq
  \sum_{n=0}^{N_{\rm KK}}
  B_n
  \log_2
  \!\left(
    1+\frac{|H_n|^2 P_n}{\mathcal{N}_n}
  \right),
  \qquad
  \sum_{n=0}^{N_{\rm KK}}P_n\leq P_{\rm tot}.
  \label{eq:parallelcapacity_bound}
\end{equation}
Here \(P_n\) is the power allocated to the \(n\)-th resolved channel,
\(\mathcal{N}_n\) is the noise power integrated over the same bin, and
\(B_n\) is the effective bandwidth assigned to that channel. Equality
is obtained only in the ideal resolved, diagonal, Gaussian limit. The
inequality sign emphasises that Eq.~\eqref{eq:parallelcapacity_bound}
is an upper bound in the physical inter-brane setting.

The effective bin width is set by the largest of the physical
broadening, the inverse observation time, and the detector frequency
resolution,
\begin{equation}
  B_n
  \sim
  \max\!\left\{
    \Gamma_n^{\rm eff},\;
    T_{\rm obs}^{-1},\;
    \Delta\nu_{\rm det}
  \right\}.
  \label{eq:effectivebandwidth}
\end{equation}
Here \(\Gamma_n^{\rm eff}\) denotes the effective linewidth expressed
as an ordinary frequency, including any physical broadening induced by
localised interactions, ultraviolet effects, warping, curvature,
finite brane thickness, or other deviations from the ideal flat
compactification. In the strict lossless idealisation,
\(\Gamma_n^{\rm eff}=0\), and the bin width is determined by finite
observation time and detector resolution. With this convention,
\(P_n\) and \(\mathcal{N}_n\) must be interpreted as signal and noise
powers integrated over the same frequency bin, so that the argument of
the logarithm is dimensionless.

\subsection*{Multiplexing versus power}

The gain associated with the KK tower is not a gain in the
gravitational coupling of a single resolved mode. That coupling remains
Planck-suppressed in the sense discussed in
Section~\ref{sec:framework}. The gain is a multiplexing gain: above
threshold, the number of available propagating modes grows. For the
flat interval,
\begin{equation}
  N_{\rm KK}(\omega)
  \sim
  \frac{\omega L}{\pi},
\end{equation}
up to the ultraviolet cutoff encoded by the form factors \(F_n\).

The candidate transmitters of Section~\ref{sec:emitters} occupy
different parts of this trade-off. Compact-object mergers provide
large gravitational luminosity but remain below the first KK
threshold for the benchmark compactification. Black-hole evaporation
may populate many KK modes thermally, with model-dependent branching
into the bulk. Photon--graviton conversion is coherent and
spectrally selective, but has negligible per-mode conversion
probability.

The robust conclusion is structural. The KK tower turns the
inter-brane link into a geometry-defined linear channel with a
potentially large number of propagating modes. A numerical capacity,
however, is not a property of the geometry alone. It depends on the
source, the detector, the noise covariance, and the degree to which the
KK modes are physically resolvable.

\section{Encoding Information in the Kaluza--Klein Tower}
\label{sec:encoding}

The preceding section described the inter-brane link as a linear
channel. We now discuss how information may be encoded when more than
one KK mode is available. The central point is that the KK tower does
not merely add final states. It enlarges the space of possible
codewords. Information may be encoded in amplitudes, phases,
polarisations, temporal structure, or in the occupation pattern of KK
levels.

\subsection*{KK-space codewords}

In a resolved KK regime, a transmitted symbol can be represented as a
vector
\begin{equation}
  \mathbf{s}
  =
  (s_0,s_1,\ldots,s_{N_{\rm KK}})
  \in
  \mathbb{C}^{N_{\rm KK}+1},
  \label{eq:KKcodeword}
\end{equation}
where \(s_n\) is the complex signal amplitude assigned to the \(n\)-th
mode. Binary occupancy encoding \(s_n\in\{0,1\}\) is the simplest
case, but more general codewords may use continuous amplitudes,
relative phases, correlated excitations, or time-dependent
modulation.

The receiver output is
\begin{equation}
  \mathbf{r}
  =
  H\,\mathbf{s}
  +
  \mathbf{n},
  \label{eq:MIMO}
\end{equation}
where \(H\) and \(\mathbf{n}\) are defined in
Section~\ref{sec:capacity}. For the flat interval in the resolved
limit,
\begin{equation}
\begin{aligned}
  H_{mn}(\omega)
  &\simeq
  \delta_{mn}\,H_n(\omega),
  \\
  H_n(\omega,\rho)
  &\propto
  f_n(y_A)f_n(y_B)\,
  F_n\,
  \frac{e^{ik_n \rho}}{4\pi \rho}.
\end{aligned}
  \label{eq:Hn_resolved}
\end{equation}
where \(\rho\) is the ordinary three-dimensional propagation distance
along the brane. Tensor, resonance, and detector-response factors are
included in \(H_n\). In a general compactification, or for a detector
that does not diagonalise the KK basis, \(H\) may be non-diagonal.

\subsection*{Optimal communication basis}

The basis that maximises distinguishability at fixed power is
determined by the eigendecomposition
\begin{equation}
  H^\dagger\Sigma^{-1}H
  =
  U\Lambda U^\dagger.
  \label{eq:diag}
\end{equation}
The columns of \(U\) are the optimal input modes, while the eigenvalues
in \(\Lambda\) give their effective gains. For the flat interval with a
diagonal noise covariance and a detector aligned with the KK basis,
\(U=I\) and the KK modes themselves are optimal. In a generic
compactification, the optimal codewords are linear combinations of KK
states determined jointly by geometry and detector noise.

The compactification therefore acts as a coding matrix in a sense
analogous to MIMO precoding. The difference is that here the matrix is
not engineered directly: it is fixed by the spectrum, eigenfunctions,
brane locations, and detector projection.

\subsection*{Water-filling and effective rank}

In the diagonal resolved limit, maximising
Eq.~\eqref{eq:parallelcapacity_bound} over \(\{P_n\}\) gives the usual
water-filling solution,
\begin{equation}
  P_n
  =
  \left(
    \mu-\frac{\mathcal{N}_n}{|H_n|^2}
  \right)_+,
  \label{eq:waterfilling}
\end{equation}
where \(\mu\) is the water level, equivalently the Lagrange multiplier
associated with the total power constraint, fixed by
\(\sum_n P_n=P_{\rm tot}\). Only modes satisfying
\(\mu>\mathcal{N}_n/|H_n|^2\) receive nonzero power. The number of such
modes defines the effective rank,
\begin{equation}
  r_{\rm eff}
  =
  \#\!\left\{
    n:\mu>\frac{\mathcal{N}_n}{|H_n|^2}
  \right\}
  \leq
  N_{\rm KK}.
  \label{eq:reff}
\end{equation}
This is the relevant measure of usable communication dimensionality.
It can be much smaller than the number of kinematically open KK modes.
For the flat interval, the endpoint overlap
\(|f_n(y_A)f_n(y_B)|^2\) is mode-independent for \(n\geq1\). The
remaining mode dependence of \(|H_n|^2\) arises from propagation near
threshold, the phase structure, the form factor \(F_n\), tensor
projections, and the detector response. In a warped or otherwise
non-uniform compactification, the overlap factors themselves can vary
strongly with \(n\), making the geometry a direct participant in the
water-filling optimisation.

\subsection*{Codebooks and decoding metric}

A codebook is a finite set of allowed transmitted vectors,
\begin{equation}
  \mathcal{C}
  =
  \{
    \mathbf{s}^{(1)},\ldots,\mathbf{s}^{(M)}
  \}.
  \label{eq:codebook}
\end{equation}
For binary occupancy encoding, the Hamming distance between input
patterns is
\begin{equation}
  d_H(\mathbf{s},\mathbf{s}')
  =
  \sum_n |s_n-s_n'|.
  \label{eq:Hamming}
\end{equation}
The physically relevant distance for maximum-likelihood decoding under
Gaussian noise is instead the channel-weighted distance in received
signal space,
\begin{equation}
  d^2(\mathbf{s},\mathbf{s}')
  =
  (\mathbf{s}-\mathbf{s}')^\dagger
  H^\dagger\Sigma^{-1}H
  (\mathbf{s}-\mathbf{s}').
  \label{eq:Mahalanobis}
\end{equation}
This is the Mahalanobis distance induced by the channel and the noise.
It controls the probability of confusing two codewords and sets the
minimum-distance requirements for reliable decoding. The
compactification geometry enters through \(H\), thereby inducing an
effective inner product on codeword space.

\subsection*{Sparse occupancy codes}

The counting of sparse occupancy patterns is a standard
combinatorial ingredient in coding theory~\cite{CoverThomas2006}.
In the sparse regime \(k\ll r_{\rm eff}\), information is encoded
primarily in the identity of the \(k\) occupied modes rather than in
their amplitudes. Before imposing distance constraints, the number of
possible occupancy patterns is
\begin{equation}
  N_{\rm patt}
  =
  \binom{r_{\rm eff}}{k}.
  \label{eq:Npatt}
\end{equation}
For \(k\ll r_{\rm eff}\),
\begin{equation}
  \log_2 N_{\rm patt}
  =
  \log_2\binom{r_{\rm eff}}{k}
  =
  k\log_2\!\left(\frac{e\,r_{\rm eff}}{k}\right)
  +
  O(\log k).
  \label{eq:sparsecode}
\end{equation}
This is only a combinatorial upper bound on the codebook size. A
usable codebook must also satisfy the minimum-distance condition set
by Eq.~\eqref{eq:Mahalanobis} and the physical power constraint.

If \(E_{\rm sym}\) denotes the energy per transmitted symbol, the
ideal sparse-code information per unit energy is bounded
parametrically by
\begin{equation}
  \eta
  =
  \frac{\log_2 N_{\rm patt}}{E_{\rm sym}}
  \simeq
  \frac{k}{E_{\rm sym}}
  \log_2\!\left(\frac{r_{\rm eff}}{k}\right),
  \qquad
  r_{\rm eff}/k\gg1.
  \label{eq:spectraleff}
\end{equation}
Increasing the number of KK modes improves this figure only if the
additional modes survive the water-filling threshold, are spectrally
distinguishable, and can be populated by the transmitter.

\subsection*{Geometry and asymptotic code-space growth}

For a \(d\)-dimensional compact manifold \(\mathcal{M}_d\), Weyl's law
gives the asymptotic growth of the number of Laplacian eigenmodes
below an internal momentum scale \(E\),
\begin{equation}
  N(E)
  \sim
  C_d\,\mathrm{Vol}(\mathcal{M}_d)\,E^d,
  \label{eq:Weyl}
\end{equation}
up to subleading boundary and curvature corrections, where
\(C_d\) is a dimension-dependent numerical constant. Equivalently,
for a smooth compact manifold without boundary,
\begin{equation}
 C_d
=
\frac{\omega_d}{(2\pi)^d},
\qquad
\omega_d
=
\frac{\pi^{d/2}}{\Gamma(d/2+1)},
\end{equation}
with \(\omega_d\) the volume of the unit ball in \(\mathbb{R}^d\).
Only the scaling \(N(E)\propto E^d\) will be used below.

In an optimistic regime where a finite fraction of these modes remains
above the water-filling threshold,
\begin{equation}
  r_{\rm eff}(E)
  \propto
  N(E)
  \propto
  E^d .
  \label{eq:reffweyl}
\end{equation}
For fixed sparsity \(k\) and \(r_{\rm eff}\gg k\), the sparse codebook
then satisfies
\begin{equation}
  \log_2 N_{\rm patt}
  \sim
  k\,d\,\log_2 E .
  \label{eq:geometrycoding}
\end{equation}
Thus the number of extra dimensions can appear directly in the
asymptotic growth of the code space. This is a scaling estimate, not
an achievable-rate theorem. If higher KK modes are suppressed by
form factors, weak overlaps, increasing noise, or detector
limitations, then \(r_{\rm eff}\) may grow more slowly than the raw
spectral count.

\subsection*{Channel tomography and inverse spectral geometry}

A receiver may attempt to estimate the channel using a calibration
signal. In the KK setting this means reconstructing the accessible
masses, overlap factors, and channel matrix,
\begin{equation}
  \{m_n,\ f_n(Y_A)f_n(Y_B),\ H_{mn}\}.
  \label{eq:tomographydata}
\end{equation}
where \(Y_A\) and \(Y_B\) denote the locations of the two branes in
the compact space. This is the analogue of channel estimation in
conventional MIMO communication theory. It also has a geometric
interpretation: recovering the compactification from these data is an
inverse spectral problem related to Kac's question of whether one can
hear the shape of a drum~\cite{Kac:1966}.

The inverse problem is generally ill-posed. Distinct compact
Riemannian manifolds can share the same Laplacian spectrum without
being isometric~\cite{Milnor:1964}. In the present context, the
measured set \(\{m_n,f_n(Y_A)f_n(Y_B)\}\) constrains the
compactification but does not determine it uniquely. The information
is nevertheless non-trivial: it fixes the accessible spectrum,
constrains the brane-to-brane overlaps, and may distinguish broad
classes of compactification geometries.

The compactification therefore acts simultaneously as propagation
medium, mode generator, and coding resource. Its geometry determines
the available KK spectrum, the channel matrix \(H\), the optimal
communication basis through Eq.~\eqref{eq:diag}, and the decoding
metric Eq.~\eqref{eq:Mahalanobis}. In this sense the geometry of the
extra dimensions does not merely affect the channel: it defines it.



\section{Discussion and Conclusions}
\label{sec:conclusions}

We have analysed the possibility of gravitational communication
between two otherwise isolated branes embedded in a higher-dimensional
spacetime. The setup is deliberately minimal: matter fields are
localised on the two branes, while gravity propagates through the
bulk. Starting from the linearised bulk graviton propagator, we
identified the brane-to-brane retarded Green function as the physical
propagation kernel of the inter-brane link. In the simplest flat
interval compactification, this kernel decomposes into a massless
zero mode and a tower of massive Kaluza--Klein modes with masses
$m_n=n\pi/L$ and endpoint overlap factors
$f_n(y_A)f_n(y_B)$.

The central result is structural. Below the first KK threshold, the
long-distance channel is effectively four-dimensional and is carried
only by the massless graviton. Above threshold, additional massive
KK modes become propagating. Each open mode carries its own phase,
group velocity, attenuation, and overlap factor. The KK tower
therefore transforms the gravitational link from a single propagation
channel into a geometry-defined multi-mode channel. This does not
remove the weakness of gravity: a single resolved KK graviton remains
Planck-suppressed in the same parametric sense as the zero mode.
The gain supplied by the tower is instead a multiplexing gain, through
the increase in the number of distinguishable propagation modes.

We also clarified the distinction between propagation, power transfer,
and communication capacity. The retarded Green function transfers
field amplitudes. Power transfer requires additional information about
the source, the emitted spectrum, detector response, and bandwidth.
Similarly, an information-theoretic capacity is not determined by the
compactification geometry alone. It depends on the projected channel
matrix, the input covariance, the detector noise covariance, the
available bandwidth, and the degree to which the KK modes are
physically resolvable. In the resolved limit, the tower can be treated
as a set of approximately parallel subchannels, leading to the usual
water-filling structure. Away from this limit, the problem is a
genuine MIMO channel whose optimal input basis is determined by
$H^\dagger\Sigma^{-1}H$.

The candidate transmitters discussed in
Section~\ref{sec:emitters} illustrate complementary regimes rather
than realistic engineering proposals. Compact-object mergers provide
enormous gravitational luminosity, but their frequencies are far below
the first KK threshold for the benchmark compactification scale
considered here. Low-mass black holes can act as broadband thermal
emitters and may populate many KK modes if their Hawking temperature
exceeds the relevant thresholds, but their emission is stochastic and
their branching into bulk modes is model-dependent. Artificial
black-hole bursts are useful only as a limiting thought experiment
involving extreme energy concentration. Photon--graviton conversion
provides the opposite limiting case: coherent, tunable, and
mode-selective excitation, but with an extremely small per-mode
conversion probability. These examples emphasise the trade-off between
power and control.

\medskip
\noindent\textbf{Assumptions and limitations.}
The analysis rests on assumptions that are speculative. We assume the
existence of a second brane supporting localised observers, and we
assume that some source on our brane can populate bulk gravitational
degrees of freedom at frequencies high enough to access the KK sector.
Neither assumption is experimentally established. We have also worked
mostly within an idealised flat interval compactification. In a more
realistic model, warping, curvature, finite brane thickness,
brane-localised kinetic terms, and ultraviolet physics can modify the
spectrum, overlaps, widths, and form factors of the KK modes.

For this reason, the results should not be read as estimates of an
achievable communication rate. They identify the structure of the
channel, not the performance of a technological implementation. A
numerical rate would require a concrete compactification, a source
model, a detector model, and a noise model. It would also require the
calculation of KK graviton production rates and detector response
functions, both of which are strongly model-dependent. The purpose of
the present work is more modest: to show how the KK tower enters the
communication problem and to formulate the corresponding channel in a
language that makes contact with information theory.

\medskip
\noindent\textbf{Geometry as a communication resource.}
The most distinctive feature of the construction is that the compact
geometry itself becomes part of the communication system. The masses
$m_n$ determine the thresholds at which new radiative channels open.
The overlap factors determine how strongly each mode connects the two
branes. The form factors determine the range of validity of the
effective KK description. After projection onto transmitter and
receiver bases, these ingredients define the channel matrix \(H\).
Thus the compactification acts simultaneously as propagation medium,
mode generator, and coding matrix.

This observation also gives the problem an inverse-spectral aspect.
A receiver capable of resolving the KK structure could, in principle,
infer partial information about the compact space from the measured
set of masses, overlaps, and channel gains. Such data would not
uniquely reconstruct the geometry: inverse spectral problems are
generally non-unique. Nevertheless, the channel response would
constrain the compactification and could reveal geometric information
before any deliberate message is decoded.

\medskip
\noindent\textbf{Outlook.}
Several directions are natural. The first is to replace the flat
interval by explicit compactifications with nontrivial geometry,
warping, or brane-localised terms, and to compute the corresponding
spectra and overlap factors. The second is to analyse concrete source
models, including their branching fractions into bulk gravitational
modes and their coherence properties. The third is to develop detector
models capable of projecting onto the relevant KK structure, including
the associated noise covariance. Finally, one can study the channel
tomography problem itself: how much geometric information about the
bulk can be recovered from a finite set of measured KK thresholds and
overlaps?

The picture that emerges is therefore not a claim of practical
feasibility, but a change of perspective. If extra dimensions and
neighbouring branes exist, another world need not be separated from us
by astronomical distance. It may be separated by a microscopic
displacement in the compact space, while remaining effectively hidden
because the only shared interaction is gravity. The difficulty is then
not distance, but coupling. In that setting, the Kaluza--Klein tower is
more than a spectrum of gravitational excitations: it is the natural
set of modes through which two otherwise isolated worlds could, in
principle, exchange information.


\section*{Acknowledgements}

I would like to thank Yifan Chen for valuable discussions. I would also
like to thank TDLI for its hospitality during the preparation of this
work.

%

\end{document}